# Metasurface-Integrated Polarization-Insensitive LCoS for Projection Displays

Xiangnian Ou[1,#], Yueqiang Hu[1,2,3,#,*], Dian Yu[1], Shulin Liu[1], Shaozhen Lou[1], Zhiwen Shu[2], Wenzhi Wei[1], Man Liu[1], Ping Yu[4], Na Liu[5], and Huigao Duan[1,2*]

[1] National Research Center for High-Efficiency Grinding, College of Mechanical and Vehicle Engineering, Hunan University, Changsha 410082, P.R. China

[2] Greater Bay Area Institute for Innovation, Hunan University, Guangzhou 511300, P.R. China

[3] Advanced Manufacturing Laboratory of Micro-Nano Optical Devices, Shenzhen Research Institute, Hunan University, Shenzhen 518000, P.R. China

[4] Songshan Lake Materials Laboratory, Dongguan 523808, P.R. China

[5] 2nd Physics Institute, University of Stuttgart, Pfaffenwaldring 57, 70569 Stuttgart, Germany

[#]These authors contributed equally to this work.

[*]Corresponding authors. Email: huyq@hnu.edu.cn, duanhg@hnu.edu.cn

**Abstract**

Liquid crystal on silicon (LCoS) panels, renowned for their high resolution and fill-factor, are integral to modern projection displays. However, their inherent polarization sensitivity constrains the upper limit of light utilization, increases system complexity and restricts broader applicability. Here, we demonstrate a dual-layer metasurface-integrated LCoS prototype that achieves polarization-insensitive, addressable amplitude modulation in the visible. Polarization sensitivity is eliminated in the reflective architecture through polarization conversion in the underlying metasurface and polarization-sensitive phase modulation of the liquid crystals (LC). This is further enhanced by the electrically tunable subwavelength grating formed by the upper metasurface and LC, resulting in a high-contrast, polarization-insensitive optical switch. We showcase a 64-pixel 2D addressable prototype capable of generating diverse projection patterns with high contrast. Compatible with existing LCoS processes, our metasurface device reduces system size and enhances energy efficiency, offering applications in projectors and AR/VR displays, with the potential to redefine projection chip technology.

## Introduction

With advancing technology and growing market demand, projection display technology is trending towards miniaturization, high resolution, high contrast, and low energy consumption. Current mature light engines for projection displays mainly include liquid crystal display (LCD), digital light processing (DLP), and liquid crystal on silicon (LCoS)[1]. LCDs modulate light amplitude by twisting the incident linear polarization by liquid crystal (LC) molecules under voltage and with a polarizer to construct images, making them popular in entry-level projection displays despite lower contrast and larger pixel sizes[2]. DLP systems use digital micromirror devices (DMDs), consisting of MEMS-based flip-flop micro-mirrors as illustrated in Fig. 1a, to control grayscale by switching time-duty cycles[3]. The system offers high contrast without the need for polarized incident light. However, its resolution is constrained, and the manufacturing costs are relatively high. LCoS, a reflective technology combining LCD with mature silicon integrated circuit technology, offers smaller pixel sizes with higher resolution and fill factor, which positions it well for future 8K and beyond displays[4]. However, LCoS's reliance on linearly polarized light necessitates polarizing optics in Fig. 1b, thereby increasing system complexity. When using common non-polarized light sources like LEDs, the light utilization efficiency is capped at 50%. Additionally, the linear polarization output limits adaptability in applications like heads-up displays (HUD), where polarized glasses can block the projected images[5]. It also struggles to adapt to multi-polarization scenarios, such as orthogonal polarization 3D displays[6] or multi-focus systems with varying polarization states[7]. Two strategies have been proposed to address these challenges: using polarization-insensitive LC materials[8, 9], which require higher drive voltages and advanced packaging, or modifying the LCoS structure by integrating thin-film wave plates[10], which increases drive voltage due to added material thickness. Both approaches face significant technical hurdles.

In this work, we propose and experimentally demonstrate a polarization-insensitive metasurface-integrated LCoS (meta-LCoS) prototype for projection displays in the visible spectrum as illustrated in Fig. 1c. Metasurfaces, which are two-dimensional artificial materials arrayed with sub-wavelength-sized scatterers, have emerged as a transformative platform for light modulation[11-16]. Previously, the introduction of various mechanisms to impart tunability to metasurfaces for realizing novel multifunctional dynamic devices has garnered widespread interest[17-22]. However, there has been little consideration of how metasurfaces can enhance the capabilities of mature optical modulation devices[23]. Here the metasurfaces are integrated onto the aluminum (Al) electrode pixels of the LCoS panel to eliminate polarization-sensitive properties while achieving high-contrast light projection, as conceptually shown in Fig. 1d. Each pixel contains metasurface super unit cells with paired nanorods overlaid with alternating subwavelength nanograting and LC. The combination of polarization conversion in the metasurface and polarization-sensitive phase modulation in the LC allows the device to function with unpolarized light. An initial π-phase difference between the two columns creates destructive interference for the "off" state, while voltage control of the LC modulates the phase difference to enable high-contrast switching between "on" and "off". By pixelating the two-dimensional electrodes, a 64-pixel addressable 2D prototype chip is achieved, enabling the dynamic generation and projection of programmable intensity images. Fully compatible with existing LCoS fabrication processes, this device reduces system complexity, enhances energy efficiency, and lowers production costs, offering promising applications in projectors and AR/VR microdisplays and setting the stage for the next generation of projection chip technologies.

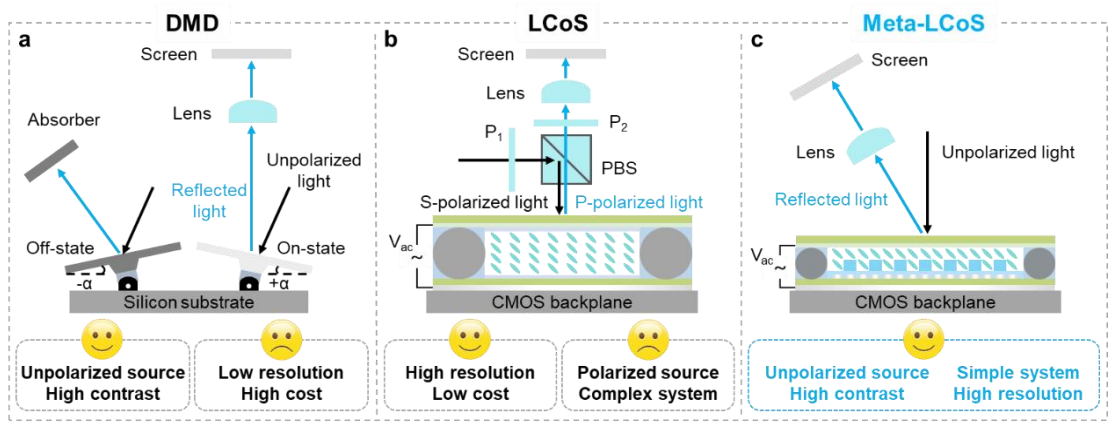

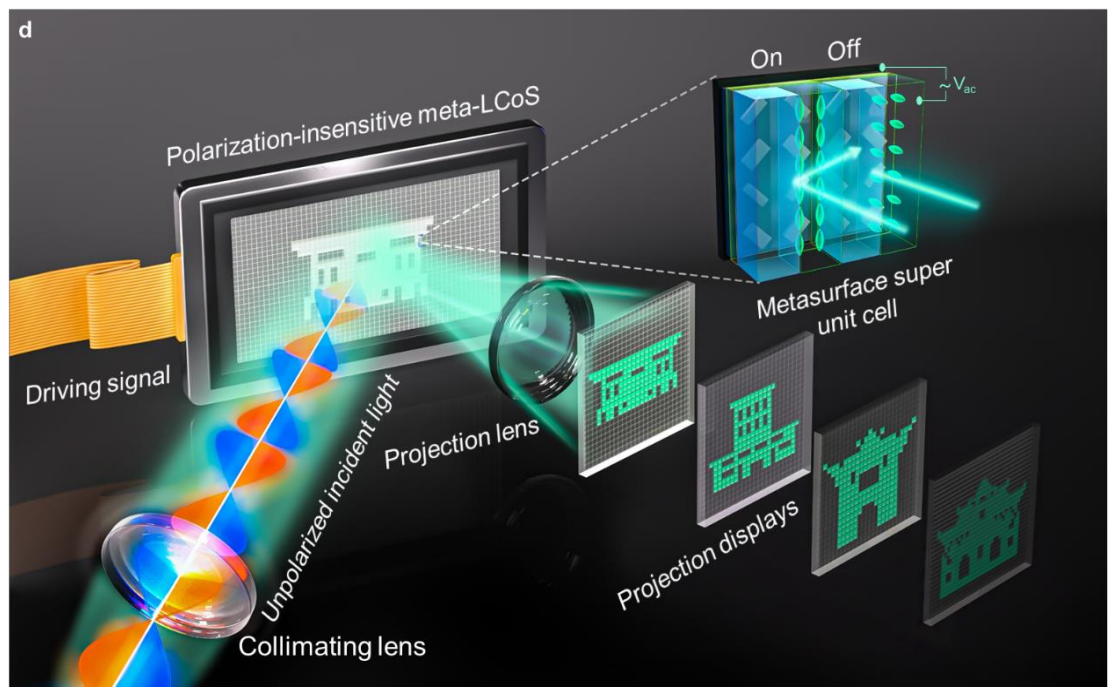

**Fig. 1. Schematics of the projection principles of light engines and the proposed metasurface-integrated polarization-insensitive LCoS projection display. a-c**, Schematic diagrams of light engine DMD (**a**), LCoS (**b**) and the proposed polarization-insensitive meta-LCoS (**c**). The lens serves as the projection lens for focusing and magnifying the images; $P_1$ and $P_2$ represent polarizers, while PBS is a polarizing beam splitter cube. With meticulous design, the proposed meta-LCoS enables high-contrast optical switching in unpolarized light, eliminating the need for bulky polarizing optics and thereby simplifying the projection system. Furthermore, the proposed meta-LCoS achieves smaller pixel sizes and higher resolution due to the utilization of CMOS process. **d**, Schematic of the proposed polarization-insensitive meta-LCoS projection display. The programmable images, driven by different digital signals, can be dynamically displayed in the far field under unpolarized incident light.

## Results

### Mechanism and design of the polarization-insensitive optical switching

Figure 2a illustrates the detailed structure of a single super unit cell and the principle of eliminating the polarization sensitivity. The super unit cell consists of two columns of Al nanorods on the reflective panel, with a dielectric nanograting and a LC layer of thickness $d$ alternately arranged on top. The LC, a typical birefringent material, exhibits polarization-sensitive refractive index responses, namely the extraordinary and ordinary refractive index ($n_e$ and $n_o$). In the figure, the LC molecules rotate within the $yz$-plane under a certain voltage, with a tilt angle $\theta_{LC}$ relative to the $y$-axis. To achieve polarization-insensitive modulation in LC, a combination of polarization conversion of the underlying metasurface and polarization-sensitive phase modulation of the LC is implemented. The resonant phase of asymmetric Al nanorods is utilized to form half-wave plates, with each nanorod's rotation angle set to either 45° or -45° along the *x*-axis. When *x*-polarized light is incident on the LC-covered columns, an initial propagation phase, denoted as $\varphi_{p1}=n_o kd$ is established. *k* is the wave vector of the incident light. The metasurface-based half-wave plate then modulates the light and reflects it, converting the polarization to *y*-polarization, while simultaneously imprinting a phase $\varphi_m$. As the light passes through the LC again, another propagation phase $\varphi_{p2}=n_{LC} kd$ is established. The equivalent extraordinary refractive index of the LC ($n_{LC}$) is changed from $n_e$ to $n_o$ and can be expressed as a function of the tilt angle $\theta_{LC}$ of the LC director from 0 to 90° controlled by voltage as $n_{LC}=\left(\frac{\cos^2\theta_{LC}}{n_e^2}+\frac{\sin^2\theta_{LC}}{n_o^2}\right)^{-1/2}$. Therefore, the overall phase modulation of the reflected light can be written as $\varphi=\varphi_m+\varphi_{p1}+\varphi_{p2}$. When *y*-polarized light is incident, the phase modulations during incidence and reflection are swapped compared to *x*-polarized light, but the total phase modulation remains the same, as shown in Fig. 2a. This ensures consistent phase modulation for all polarized incidence, which can be described by *x*- and *y*-polarized bases, thereby eliminating polarization sensitivity in the LC. For nanograting-covered columns, polarization-insensitive phase modulation can be naturally achieved with isotropic materials.

This polarization-insensitive property of the LC-covered columns can be further explained in terms of the transmission matrix which can be expressed as

$$T=\begin{pmatrix} e^{i\varphi_{p1}} & 0 \\ 0 & e^{i\varphi_{p2}} \end{pmatrix} R(45°) \begin{pmatrix} e^{i\varphi_m} & 0 \\ 0 & e^{i(\varphi_m+\pi)} \end{pmatrix} R(-45°) \begin{pmatrix} e^{i\varphi_{p1}} & 0 \\ 0 & e^{i\varphi_{p2}} \end{pmatrix}$$
$$=e^{i(\varphi_m+\varphi_{p1}+\varphi_{p2})} \begin{pmatrix} 0 & 1 \\ 1 & 0 \end{pmatrix} \quad (1)$$

where $R(\alpha)=\begin{pmatrix} \cos\alpha & -\sin\alpha \\ \sin\alpha & \cos\alpha \end{pmatrix}$ is the rotation matrix. The transmission matrix demonstrates that the incident linearly polarized light is converted to orthogonally polarized light after modulation and reflection by the device with an accompanying polarization-insensitive modulation phase, which is consistent with the previous analysis.

Eliminating polarization sensitivity means that traditional polarization-selective methods for amplitude modulation (i.e. optical switching) are no longer applicable. To address this, we designed a dynamic subwavelength grating structure, formed by the upper nanograting and LC, to achieve high-contrast amplitude modulation. Figure 2b illustrates the specific structure design of the super unit cell, along with the corresponding phase profiles and resulting projection intensities during the optical switching. A phase difference, $\Delta\varphi_x$, exists between the two columns of the super unit cell, determined by the underlying nanorods, and the nanograting and LC layers. This phase difference can be modulated to control the transition between propagating and evanescent waves when the nanograting period is subwavelength (see also Supplementary Information Section 1). When $\Delta\varphi_x = j\pi$ ($j = 0, 1, 2, 3 \cdots$), the reflected modulated light is completely suppressed due to destructive interference between neighboring columns. Conversely, as $\Delta\varphi_x$ deviates from $\pi$, the reflected light reappears, fully switching "on" when $\Delta\varphi_x = 2j\pi$. The geometric phase of the nanorods is used to create an initial phase difference, $\Delta\varphi_g=2\Delta\theta=\pi$, where $\Delta\theta$ is the angle difference between the Al nanorods, meaning that the two columns of nanorods are perpendicular to each other. This precise geometric phase difference, introduced by the nanorod rotation, stabilizes the "off" state, enabling high contrast performance. Additionally, the phase difference from the LC and nanograting during incidence and reflection under arbitrary polarized or for unpolarized light, $\Delta\varphi_p=\Delta\varphi_{p1}+\Delta\varphi_{p2}=(n_o+n_{LC}-2n_G)kd$, provides a dynamic modulation degree of freedom. By dynamically adjusting $\Delta\varphi_p$, the total phase difference $\Delta\varphi_x=\Delta\varphi_g+\Delta\varphi_p$ can be tuned from 0 to $\pi$, thus enabling the dynamic optical switching functionality.

To ensure high contrast performance by separating the modulated light from the zero-order light, a linear phase gradient ($\frac{d\phi}{dy}=\frac{2\pi}{4P}$, where $P$ is the lattice period of single unit cell which are optimized to be less than half of the wavelength of the incident light) is also introduced along the y-direction to form an off-axis reflection point as the switching point (see also Supplementary Information Section 2). The phase gradient is achieved by adjusting the geometry and rotation angle of the nanorod arrays, combining geometric and resonant phases. The angle of off-axis reflection light can be determined using the generalized Snell's law[24]. By combining the aforementioned design, we can theoretically achieve a polarization-insensitive, high-contrast, electrically driven optical switching with an off-axis reflection point, as illustrated in the two states shown in Fig. 2b.

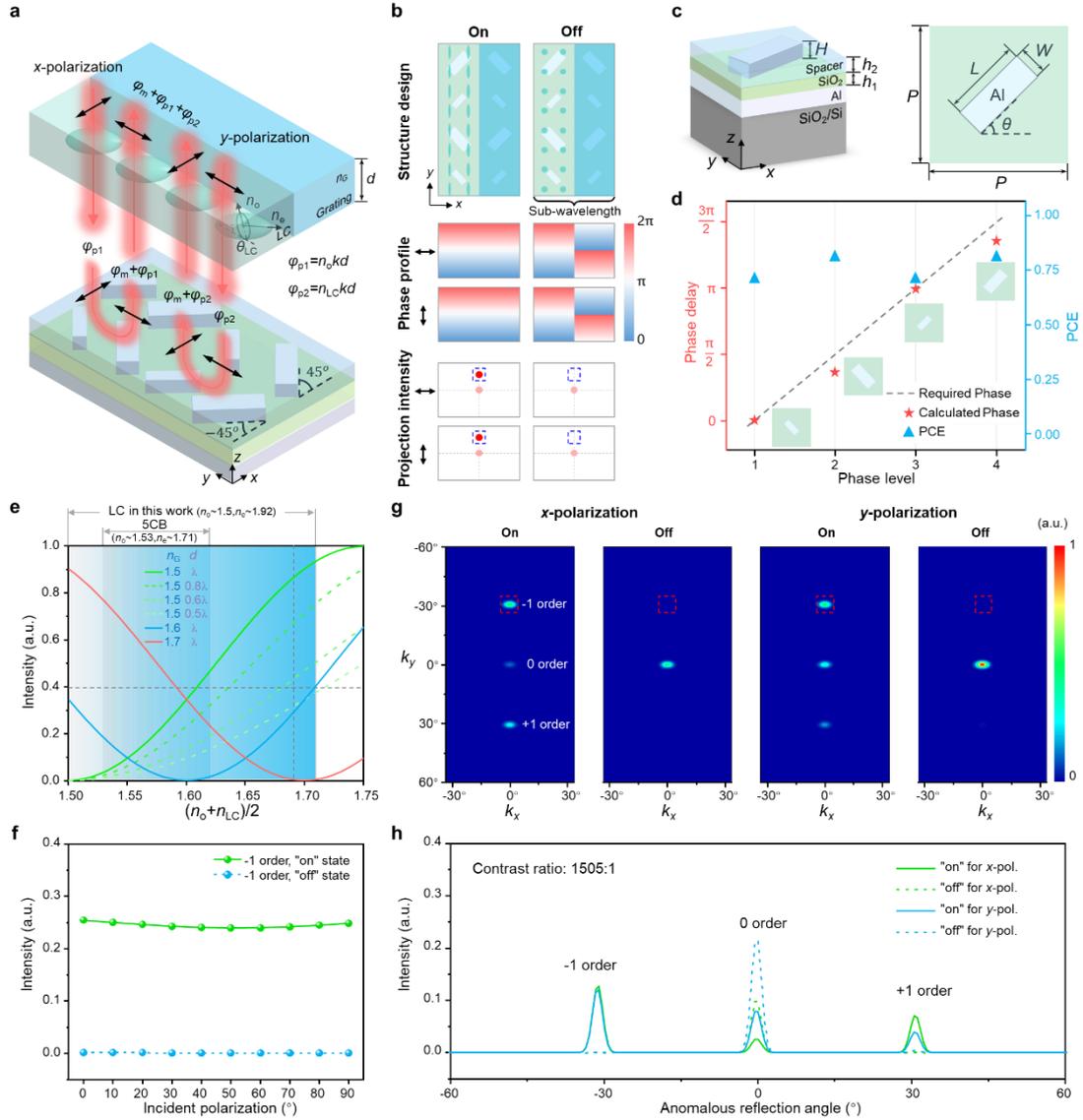

**Fig. 2. Design and Simulation of the proposed polarization-insensitive meta-LCoS. a**, The detailed structure of a single super unit cell and the principle of eliminating the polarization sensitivity of the polarization-insensitive meta-LCoS. **b**, Schematic of the designed super unit cell, along with the corresponding phase profiles and the resulting projection intensities. **c**, The three-dimensional view and top view of the metasurface unit cell of the designed Al-SiO$_2$-Al MIM structure. The unit cells form a square lattice with a period of $P$ = 260 nm at the operational wavelength of 532 nm. The thickness of each layer at the three operational wavelengths was optimized to the same parameters (100 nm for the Al mirror, $h_1$ = 50 nm for the SiO$_2$, $h_2$ = 70 nm

for the spacer layer and *H* = 50 nm for the nanorods) for the purpose of facilitating fabrication. **d**, Phase delay and polarization conversion efficiency for the different phase levels. Four phase levels are used in the design. The stars indicate the nanorods selected to construct the metasurface at the operational wavelength of 532 nm. **e**, Calculated intensity of the off-axis reflection in response to changes in the refractive index and height of the LC and nanograting. **f**, Simulated normalized intensity of the off-axis reflection point at different incident polarization angle from super unit cell with operational wavelength of 532 nm. **g** and **h**, Far-field simulation results (**g**) and the normalized intensity simulation curve (**h**) from polarization-insensitive meta-LCoS at the operational wavelength of 532 nm with an off-axis reflection angle of 30.8° in the *y*-direction.

**Numerical simulation and optimization**

Figure 2c illustrates the composition of the underlying metasurface unit cell, which is designed as a metal-insulator-metal (MIM) structure of Al-SiO$_2$-Al. To prevent the infiltration of LC into the nanorods, the nanorods are coated with a spacer. Based on the proposed design methodology, the individual metasurface unit cell is simulated numerically to determine the optimal dimensions at three operational wavelengths of 465 nm, 532 nm and 633 nm (see Supplementary Information Section 3 for the detailed simulation setup). Phase control ($|\varphi_{me}-\varphi_{mo}|=\pi$) achieved through customized nanorod size leads to the highest polarization conversion efficiency (PCE). $\varphi_{me}$ and $\varphi_{mo}$ are the phase delays in the slow and fast axes of the nanorod, respectively. Considering the phase shift, PCE, and reflectivity of the metasurface unit cell, two types of nanorods with dimensions of 110 nm × 55 nm × 50 nm and 260 nm × 70 nm × 50 nm have been selected and arranged to form a half-wave plate with four phase levels at the operational wavelength of 532 nm, which then constitute the super unit cell, as illustrated in Fig. 2d. Each level incorporates the phase delay of 0, $\pi/2$, $\pi$, $3\pi/2$, successively. The selected nanorods closely match the required phase profile and exhibit an average PCE of approximately 75%. Simulation results at operational wavelengths of 465 nm and 633 nm are shown in the Supplementary Information Section 3.

For the design of the upper layer comprising LC and nanograting, the phase difference between the LC and nanograting, $\Delta\varphi_p = (n_o + n_{LC} - 2n_G)kd$, should be controlled to achieve dynamic optical switching. Fig. 2e illustrates the variation in the intensity of the off-axis reflection in response to changes in the refractive index and height of the LC and nanograting. The horizontal axis $(n_o + n_{LC})/2$ reflects the refractive index property of the LC and its dynamic variation as a function of the applied voltage. The thickness $d$ of the nanograting is set as $\lambda$, $0.8\lambda$, $0.6\lambda$ and $0.5\lambda$ accordingly, while the refractive index of the nanograting $n_G$ is varied to different values of 1.5, 1.6 and 1.7 accordingly. $\lambda$ is the wavelength of the incident light. As illustrated in the figure, when $n_G = (n_o + n_{LC})/2$, $\Delta\varphi_p = 0$, the off-axis reflection is completely extinguished due to the phase difference of nanorods, resulting in an "off" state critical for high-contrast projection displays. This indicates that the refractive index of the nanograting must be selected within the range between $n_o$ and $(n_o + n_e)/2$ to ensure the existence of the "off" state. Furthermore, with constant refractive index for both the nanograting and LC, super unit cells composed of nanograting of varying heights can generate off-axis reflections of differing intensities (gray vertical dashed lines), without affecting the "off" state. This suggests that nanograting height minimally impacts contrast but influences the "on" state intensity, enhancing fabrication robustness. Additionally, for LC with higher refractive index contrast, equivalent reflection intensities can be achieved with smaller nanograting heights (gray horizontal dashed lines), simplifying the fabrication process. Therefore, in this work, we opted to use a high birefringent LC material ($n_e$~1.92 and $n_o$~1.5 at 25 °C for 532 nm) instead of the commonly used LC (e.g. 5CB) and choose PMMA ($n_G$~1.5) as the dielectric nanograting. This design enables the device to be in the "on" state when no voltage is applied ($\Delta\varphi_p$=0.42$kd$) and an "off" state when voltage is applied to cause $n_{LC}$=$n_o$=$n_G$=1.5, resulting in $\Delta\varphi_p$=0.

Figure 2f depicts the simulated normalized intensity of the off-axis reflection point at different incident polarization angle from super unit cell with operational wavelength of 532 nm. The intensity of the off-axis reflection point remains substantially invariant under different polarization states of incident light, both in the "on" and "off" states of the device. This consistent performance further substantiates the polarization-insensitive characteristic of the design scheme. Figure 2g presents the far-field simulation results of off-axis reflection point generated at the operational wavelength of 532 nm with an off-axis reflection angle of 30.8° in the *y*-direction and the normalized intensity simulation curve is shown Fig. 2h. The red dashed box indicates the designed switching point. The simulation results demonstrate that the device is capable of producing a switching effect in both *x*- and *y*-polarization, with a high contrast ratio reaching 1505:1.

**Device fabrication and characterization**

To prove the concept, we fabricated three devices operating at wavelengths of 465 nm, 532 nm, and 633 nm, respectively. The devices were mainly fabricated by electron beam lithography (EBL) overlay process, thermal evaporation, lift-off and LC packing, as illustrated in Fig. 3a (see Methods for details of the process)[25, 26]. To simplify the fabrication process, the nanograting height was set at 300 nm. The long axes of LC molecules are aligned along the nanograting trenches due to the anchoring effect of the alignment layer on the glass cover and the nanograting. Figure 3b, c shows the scanning electron microscopy (SEM) images of fabricated Al nanorod arrays and the subsequent PMMA nanograting fabricated using the EBL overlay process before LC infiltration, respectively.

The optical measurement setup for characterizing the intensity and polarization insensitivity of the off-axis reflected light after LC packaging is illustrated in Fig. 3d. To ensure that the intensity of light with different linear polarization incident on the device is uniform, a polarizer and a quarter-wave plate are utilized to convert the light source into circularly polarized light. Figure 3e, f, and g illustrates the intensity of the off-axis reflected light with the applied voltage under three operating wavelengths, respectively. As the applied voltage increases, the intensity of the off-axis reflected light decreases continuously (see also Supplementary Movie 1-3). The one-to-one correspondence between voltage and intensity allows for grayscale modulation in projection displays. This phenomenon can be attributed to the gradual change in the refractive index of LC from $n_e$ to $n_o$ under voltage. The experimental observations align with the calculated result depicted in Fig. 2e. The drive voltage is primarily influenced by the intrinsic characteristics of the LC and its thickness, and can be further optimized to suit the CMOS backplane. Intensity modulations with contrast ratios of 58.6:1, 81.3:1, and 62.4:1 was achieved at 465 nm, 532 nm, and 633 nm wavelengths, respectively. The contrast ratio is lower than that predicted by the simulation results due to the refractive index of PMMA nanograting dose not precisely match the ordinary refractive index of LC. It is also affected by the near-field coupling of neighboring nanorods and the inaccuracies of the inherent fabrication process.

Figure 3h, i, and j shows the intensity of off-axis reflected light at different incident polarization angle for devices with operational wavelengths of 465 nm, 532 nm and 633 nm, respectively. In light of the inherent fluctuations in intensity of incident light with different polarization, the intensity ratio of the off-axis reflection light to the incident light ($I_{RL}/I_{IL}$) was characterized and normalized. The black curve represents the calculated modulation intensity of traditional LCoS as a function of the incident polarization angle. In comparison to the traditional LCoS device, the fabricated meta-LCoS demonstrated a reduced intensity fluctuation, and the intensity ratio of the off-axis reflection light to the incident light remained essentially constant. This experimental result verifies the polarization-insensitive property of the device. The switching time of this device, utilizing a 3 μm thick LC layer, is 40 ms and 65 ms for the "off" and "on" states, respectively, comparable to that of traditional LCoS device (see also Supplementary Information Section 4). It is noteworthy that only the LC molecules present in the PMMA trenches contribute to the amplitude modulation, thereby promoting a reduction in the LC thickness. This reduction facilitates decreased driving voltage and pixel size, enhancing the switching rate and resolution of the LCoS chip.

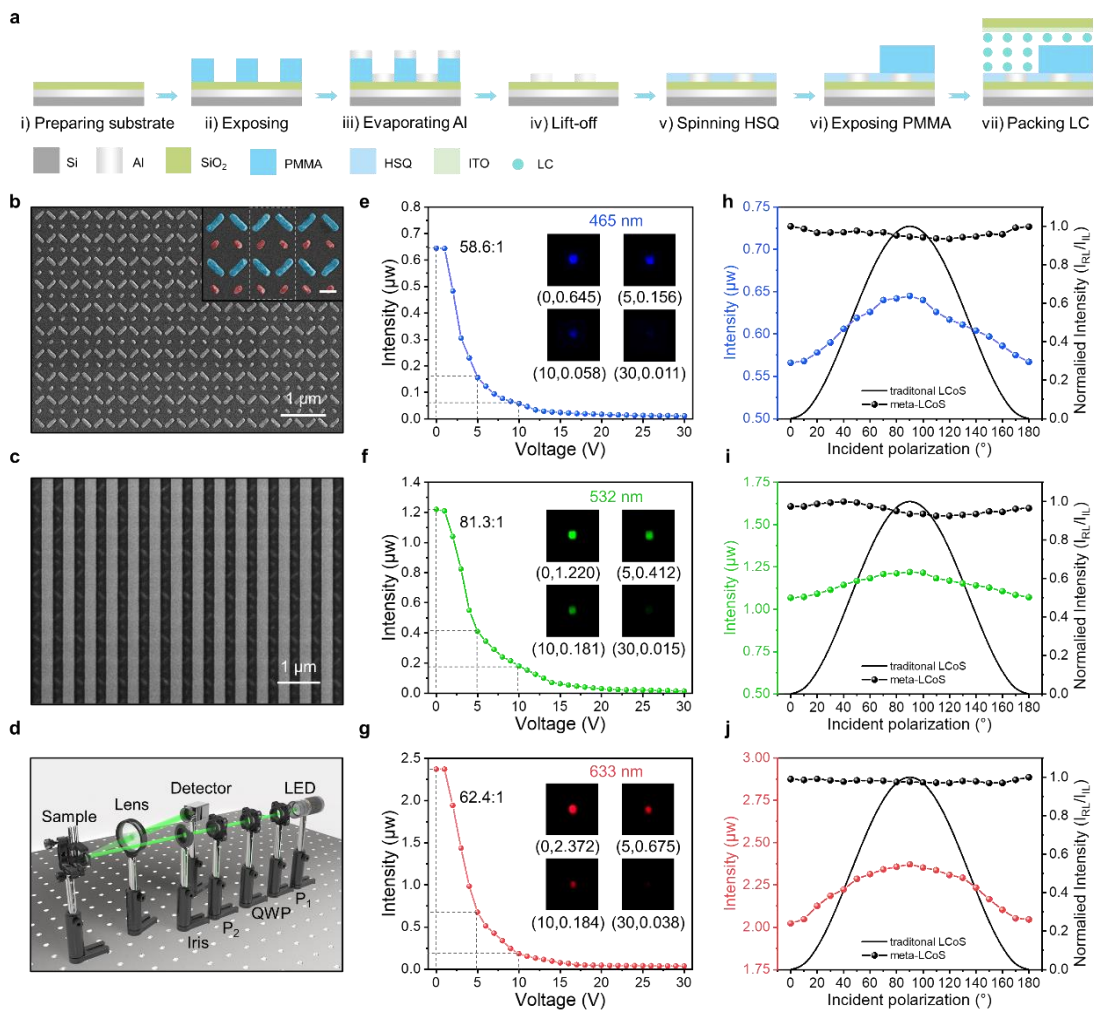

**Fig. 3. Fabrication and characterization of the proposed polarization-insensitive meta-LCoS devices. a**, Fabrication process of the polarization-insensitive meta-LCoS devices. **b, c**, SEM images

of fabricated Al nanorod arrays and devices prior to LC infiltration, respectively. Scale bar is 1 μm. Scale bar of the illustration is 200 nm. **d**, Optical measurement setup for characterizing the off-axis reflected light. $P_1$ and $P_2$ represent polarizers, while QWP is a quarter-wave plate. **e-f**, Intensity of the off-axis reflected light as a function of the applied voltage at operational wavelengths of 465 nm, 532 nm and 633 nm, respectively. **h-j**, Intensity of off-axis reflected light at different incident polarization angle for devices with operational wavelengths of 465 nm, 532 nm and 633 nm, respectively.

**64-pixel polarization-insensitive meta-LCoS dynamic projection display**

Our concept can be further extended to projection displays with multi-pixel addressing control. Specifically, we fabricated a 64-pixel 2D addressable polarization-insensitive meta-LCoS prototype device for projection display applications. Figure 4a shows a photograph of the final test device bonded to the leads of a printed circuit board (PCB). We first fabricated 64 square Al pixels, each measuring 50 μm, with a pixel pitch of 72 μm, serving as reflective mirrors, as depicted in the optical microscope photograph in Fig. 4b. The Al pixels were connected by leads and bonded to PCB contacts. Both the Al pixels and their leads were simultaneously exposed using a direct laser writing device. Subsequently, the metasurfaces were fabricated onto the pixels using the process outlined in Fig. 3a, as shown in the SEM image in Fig. 4c (see also Supplementary Information Section 5). Notably, the size and number of pixels can be further optimized using mature CMOS processes, as the generation of off-axis reflected light is solely related to the phase gradient period of the metasurface super unit cell.

Figure 4d presents an image projected onto a screen with all 64 pixels in the "on" state, captured by a camera. Figure 4e, f shows the dynamic projection of the letters "h", "n" and "u" under $x$-polarized and $y$- polarized incident light, respectively (see also Supplementary Movie 4). The polarization-insensitive property of the device is further evidenced by the display effects observed under both $x$-polarization and $y$-polarization. To expand the capability of the concept, we conducted additional experiments using a LED source without polarizing elements, involving the dynamic projection display of numbers "0-9" and arrows, as shown in Fig. 4g and h (see also Supplementary Information Section 6 for testing setup and Supplementary Movie 5, 6). For color projection display, we consider a light engine prototype employing beam-splitting dichroic prisms based on the designed polarization-insensitive meta-LCoS operating at 465 nm (blue), 532 nm (green) and 633 nm (red) wavelengths (see also Supplementary Information Section 7).

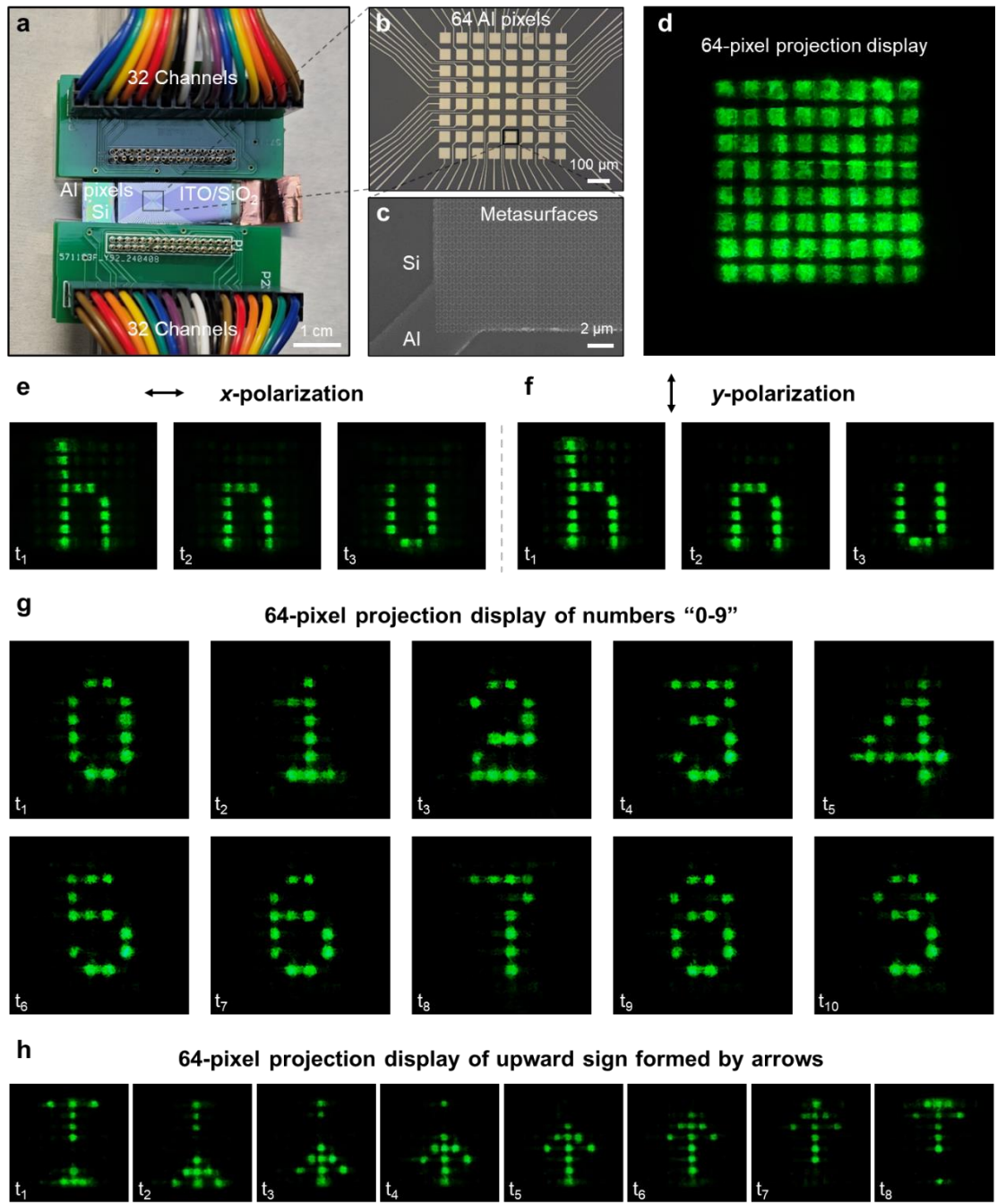

**Fig. 4. Demonstration of 64-pixel polarization-insensitive meta-LCoS dynamic projection display. a**, Photograph of the 64-pixel polarization-insensitive meta-LCoS device. Aluminum electrode leads bonded to the PCB board contacts. Scale bar is 1 cm. **b**, Optical microscope photograph of 64 Al pixels. Scale bar is 100 μm. **c**, SEM image of the metasurface fabricated onto the Al pixels using the EBL overlay technique. Scale bar is 2 μm. **d**, Projected image with 64 pixels

simultaneously in the "on" state. **e, f**, Dynamic projection display of the three letters "h", "n" and "u" in *x*-polarization and *y*-polarization, respectively. **g**, Dynamic projection display of numbers "0-9" **h**, Dynamic projection display of arrows. The image plays to form a dynamic upward sign.

## Conclusion

In summary, we have demonstrated a meta-LCoS prototype chip that achieves polarization-insensitive, addressable, high-contrast amplitude modulation at visible frequencies. The device consists of smaller super unit cells that can be independently tuned, allowing us to program the cells with voltages to produce higher resolution projection displays. Its polarization-insensitive property will also greatly simplify the configuration of LCoS projection systems. The proposed device can have a significantly reduced LC thickness due to the dynamic amplitude modulation by using only the LC molecules in the nanograting trenches. When the thickness of LC is sufficiently reduced, the grating is capable of producing uniform LC alignment without the need of any alignment layer, further simplifying device fabrication and reducing response time and drive voltage[27,28]. This design facilitates small pixel sizes and the suppression of inter-pixel crosstalk.

In addition, our concept can incorporate dielectric nanorods to enhance efficiency and contrast further, but significant challenges in fabrication must be addressed (see Supplementary Information Section 8). Moreover, other active materials with refractive index variations can also be used in the proposed design methodology for various applications[29,30]. The performance of the devices will be further enhanced by nonlocal design with consideration of near-field coupling between nanorods[31-33] and advanced inverse design methods[34-37]. Such devices will contribute to the development of next-generation customized high-resolution, high-contrast meta-LCoS optical projection devices, and will also significantly advance the field of dynamic optical metasurfaces.

## Methods

**Numerical simulation.** The responses of the device were designed and simulated using the finite-difference time-domain (FDTD, Ansys Lumerical FDTD) method. Critical cell parameters, including the length (*L*), width (*W*), thickness (*H*) and period (*P*) of the nanorods as well as the thickness of the $SiO_2$ ($h_1$) and spacer layer ($h_2$), need to be optimized to achieve proper phase, higher PCE and reflectivity. The PCE defined as the ratio of optical power of the reflected light with orthogonal polarization to the incident optical power. See Supplementary Information Section 3 for detailed simulation settings.

**Device fabrication.** The sample of single-pixel metasurfaces was fabricated as follows. A 5 nm chromium (Cr) adhesion layer and a 100 nm Al film were sputtered onto a 4-inch silicon wafer using ion beam sputtering. Next, a 50 nm silica spacer layer was sputtered. The wafer was then diced into 15×15 mm substrates. The samples were mainly fabricated using the electron beam lithography (EBL) overlay process. First, a structural layer composed of the metasurfaces and alignment marks was defined on the substrate using EBL. A 50 nm Al layer was deposited using thermal evaporation, followed by lift-off. Next, a 70 nm layer of hydrogen silsesquioxane (HSQ) was spin-coated onto the substrate and baked at 250 °C for one hour to remove the solvent. Subsequently, a 300 nm layer of PMMA resist layer was coated onto the substrate. The PMMA nanograting was fabricated by EBL and aligned with the metasurfaces of the first layer using the Al markers. The preparation of the 64-pixel samples necessitated the fabrication of Al electrode pixels as a preliminary step. Electrode patterns and alignment marks were defined on the substrate using laser direct writing (LDW) lithography on a 30 × 30 mm silicon substrate with 285 nm oxide, as shown in Fig. S5. The remaining steps are identical to those used for creating single-pixel samples, with the exception that the Al electrode pads must be carefully protected throughout the fabrication process.

**LC cell packaging.** A polyimide layer was spun on an ITO-coated glass substrate and subsequently cured on a hot plate at 100 °C for 30 min. The LC orientation directions were obtained by unidirectional rubbing of the polymer layer. After that, the glass substrate was adhered to the metasurface samples using UV-curable adhesive (NOA 81) containing 3 μm glass spacer beads. Following the application of the UV-curable adhesive, the cell was heated to 70 °C and infiltrated with LC. Once the cell has cooled to room temperature, the edges sealed with UV-curable adhesive.

# Acknowledgments

The authors thank Fan Fan and Zenghui Peng for their guidance on the LC packaging process. We acknowledge the financial support from the National Natural Science Foundation of China (Grant No. 62275078), Natural Science Foundation of

Hunan Province of China (Grant No. 2022JJ20020), the Science and Technology Innovation Program of Hunan Province (Grant No. 2023RC3101) and Shenzhen Science and Technology Program (Grant No. JCYJ20220530160405013).

**Author Contributions**

X.O. and Y.H. contributed equally to this work. Y.H., X.O. and S.Lou proposed the idea. X.O. D.Y. and S. Lou conceived and carried out the design and simulation. X.O., D.Y., Z.S., W.W. and M.L. prepared the metasurface samples. X.O. and S.Liu carried out the LC packaging. X.O., S.Liu and Y.H. conceived and performed the measurements. X.O., Y.H., P.Y., N.L. and H.D. analyzed the data and co-wrote the manuscript. Y.H. and H.D. supervised the overall project. All the authors discussed the results and commented on the manuscript.

**Competing Interests:** The authors declare that they have no competing interests.